# VR Content Capture using Aligned Smartphones

Ramanujam R Srinivasa, Joy Bose, Dipin KP
Web Services Group
Samsung R and D Institute
Bangalore, India
{rs.ramanujam, joy.bose, dipin.kp}@samsung.com

*Abstract*—There are a number of dedicated 3D capture devices in the market, but generally they are unaffordable and do not make use of existing smartphone cameras, which are generally of decent quality. Due to this, while there are several means to consume 3D or VR content, there is currently lack of means to capture 3D content, resulting in very few 3D videos being publicly available. Some mobile applications such as Camerada enable 3D or VR content capture by combining the output of two existing smartphones, but users would have to hold the cameras in their hand, making it difficult to align properly. In this paper we present the design of a system to enable 3D content capture using one or more smartphones, taking care of alignment issues so as to get optimal alignment of the smartphone cameras. We aim to keep the distance between the cameras constant and equal to the inter-pupillary distance of about 6.5 cm. Our solution is applicable for one, two and three smartphones. We have a mobile app to generate a template given the dimensions of the smartphones, camera positions and other specifications. The template can be printed by the user and cut out on 2D cardboard, similar to Google cardboard. Alternatively, it can be printed using a 3D printer. During video capture, with the smartphones aligned using our printed template, we capture videos which are then combined to get the optimal 3D content. We present the details of a small proof of concept implementation. Our solution would make it easier for people to use existing smartphones to generate 3D content.

I. INTRODUCTION

There are a number of VR and 3D devices available in the market. However, currently we are lacking in the volume of VR content. A query for available VR videos on YouTube generates vastly less search results than the number of total videos. Hence, there is currently a void for VR content generation, in contrast with VR consumption for which affordable devices like Samsung Gear VR and Google Cardboard are widely available. One of the reasons for the lack of VR content is that dedicated devices for VR content generation are relatively costly. A VR recorder today costs around $400 or $500, while Google cardboard or other VR viewers are much more affordable. This is preventing VR from reaching its potential of widespread usage.

In this paper, we present a system to align existing smartphones for VR or 3D content capture. This system will enable anyone having one or more inexpensive mobile phones with our application, to generate VR content. Given that smartphone penetration is pretty high, and decent quality smartphones with cameras can be obtained for as low as $100, this can lead to more widespread usage of VR, as long as the alignment problem is solved using our system.

Our system enables capture of 3D and/or VR content using one, two or three smartphone cameras, taking care of alignment issues so as to get optimal alignment of the cameras of the smartphones. Our objective is to keep the distance between the cameras constant and equal to the inter-pupillary distance [1] (approximately 6.5 cm as a result of the 2012 army survey [2]). One method by which we achieve this is to generate a template given the dimensions of the smartphones, camera positions and other specifications. The generated template can be printed in a 2D material (such as cardboard), or using a 3D printer. During video capture, with the media capturing devices aligned using our printed template, the devices can capture media, which are then combined to get the optimal 3D and/or VR content.

The rest of the paper is structured as follows: In section 2 we survey related work in the area of VR and 3D content generation. Section 3 outlines our approach for aligning smartphones to capture the content. Section 4 outlines the one device solution for 3D content capture, section 5 the three device solution for 360 degree videos and section 6 the two device solution. Section 7 gives details of printing a cardboard holder template to align and hold 2 smartphones to capture the content. Section 8 details modules of our app to combine the output of two smartphone video camera streams. Section 9 gives details of our prototype implementation, and section 10 concludes the paper.

II. RELATED WORK

Trenholm [3] has surveyed some existing 3D cameras, all priced at $300 or more. A number of 3D enabled mobile phones are also available [4], using autostereoscopic displays. Dual camera phones [5] are also becoming available recently, which should eventually allow VR content to be captured. But these are premium phones and VR apps are not widely available for such phones.

Current VR camera solutions include Lucidcam [6, 7], which was an Indiegogo funded project claiming to be the world's first consumer 3D VR camera. It sells for a retail price of $499 and consists of a dedicated camera with 2 lenses to capture VR. Another solution, Samsung Gear 360 [8], sells for a retail price of around $350. It is a circular device consisting of three cameras fitted at an angle of 120 degrees each.

On the other hand, devices meant for VR consumption sell relatively cheap. Google cardboard or its variants [9, 10] can be bought for around $10, and even premium VR viewers such as Gear VR [11] cost around $100.

Camerada app [12, 13] is an app for taking stereoscopic videos. It uses two viewfinders from two phones stacked over one another, syncs the phones and renders full VR video with an approximately 180-degree view. But it is difficult to hold two phones together. It does not solve the alignment problem.

The GoPro patent [14] provides a system that generates a unitary rendered image for a video from two cameras by detecting a communication coupling and syncing frames captured by the two cameras.

These related works show that it is feasible to combine 2 camera feeds for 3D content. However they do not focus on the alignment problem. They don't mention how to align smartphones to match the inter pupillary distance of around 6.5 cm.

## III. OUR APPROACH

We propose a system to create VR content using existing smartphones, aligning them with the help of a cardboard holder and combining the output of smartphone cameras to generate VR content in real time and in an inexpensive way. Our solution is both affordable and convenient.

Our approach consists of the following:

- An apparatus to align one, two or three phones in order to capture VR content
- Distance between the cameras to be equal to the inter pupillary distance (6.5 cm) to generate accurate VR output
- Software to compute the printing of the holder template on a 2D or 3D printer, given the dimensions of the mobile phones and the respective camera locations on the devices

In the following sections, we detail each of these approaches.

## IV. VR CONTENT CAPTURE WITH A SINGLE SMARTPHONE CAMERA

For capturing VR or 3D content, a minimum of two camera feeds are needed, in order to generate the depth effect. However, this can be done using a single smartphone by having a holder apparatus with a system of two mirrors, which generate the feeds that are combined into the single mobile phone camera. Similar to 3D glasses with blue and red tinted lenses, we color the mirrors red and blue, in order to generate the 3D effect. The resulting apparatus looks a bit similar to a periscope.

Hence, the apparatus for 3D content generation using a single camera has the following components:

- Two mirrors, one a double sided mirror painted blue and the other a single sided mirror painted red
- Both the mirrors are kept at the inter-pupillary distance (6.5 cm) and held by the apparatus
- Blue and red lights falling on the camera represent the left and right eye feed respectively.

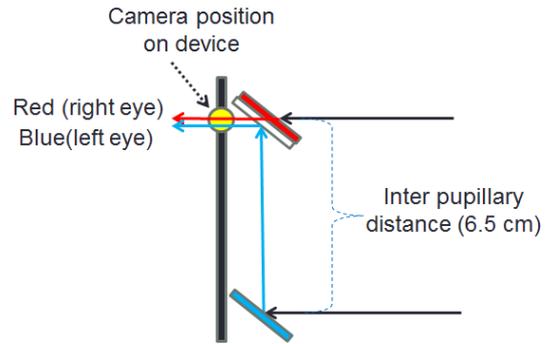

Fig. 1. Illustration for the system of 3D or VR content capture using a single smartphone camera and two mirrors, one blue and one red, tilted at an angle of 45 degrees and kept apart at the interpupilliary distance of 6.5 cm.

Fig.1. illustrates the arrangement for VR content capture with a single camera.

## V. 360 DEGREE VR CONTENT CAPTURE USING THREE SMARTPHONE CAMERAS

This replicates the functioning of Gear 360 type of devices, by using three phones. The apparatus for 360 degrees 3D content generation using three cameras has the following:

- An apparatus for three mobile phone in the shape of a equilateral triangle
- Each of the mobile phone cameras are separated by the inter-pupillary distance (6.5 cm) and held by the apparatus (cardboard or 3D printing)

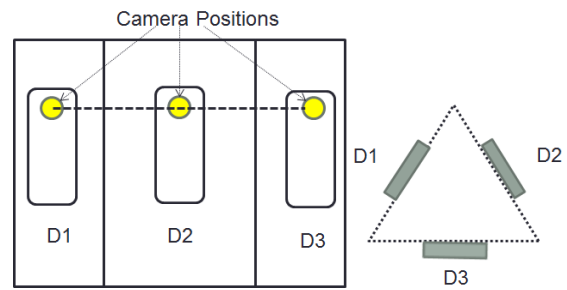

Fig. 2. Illustration for the system of 360 degrees 3D content capture using three smartphones cameras. The distance betweemn the cameras is kept equal to the inter pupillary distance. The second image shows the smartphone positioning top view when the smartphone cameras are aligned and the cardboard is folded in the form of a triangle.

Fig. 2 illustrates the arrangement for 360 degrees VR content capture with three smartphone cameras.

## VI. VR CONTENT CAPTURE WITH TWO SMARTPHONE CAMERAS

In order to capture VR content with two smartphone cameras, here too the aim is to align the smartphones such that the separation between the cameras is equal to the inter-pupillary distance of 6.5 cm.

Our system provides three types of guidance for alignment of the two smartphones:

- **Voice Guidance:** This gives voice instructions to the user to align the two devices to match the human eye spacing, for generating the VR content. The voice guidance system uses magnetometer and gyroscope sensors on the mobile devices to assist the user. It first reads the magnetic location of both the devices with respect to magnetic poles of earth on all the three axes. Then the guidance system transforms the base model as a function of magnetic poles. The devices are said to be aligned if the magnetic pole reading on all the three axis of the device matches the transformed co-ordinate model. The gyroscope is used to identify if there is a tilt in the alignment.

- **Visual Guidance:** This gives a visual grid to enable the user to position the devices one on top of the other. It transforms the base model as a function of screen dimension and pixel density. The transformed model can now be rendered as a grid on the screen. The displayed grids can be used to align the devices, by assisting the user where to place and move the two smartphones relative to each other until they are perfectly aligned. The grid generation also takes care of the orientation of the device.

- **Holder Apparatus using a template:** This enables the user to print a template on a cardboard which they can then use to pack the two devices together. It downloads a base template for each device, which contains both 3D and 2D templates. The template is split into corners and sides. The base template is morphed based on device configuration like corner radius, side width, length and depth. Once the base templates are morphed to suit each device, they are overlapped using base model. Now the template is ready to be printed on a 3D or 2D printer. The user can hold this apparatus while capturing the video simultaneously with both the smartphone cameras.

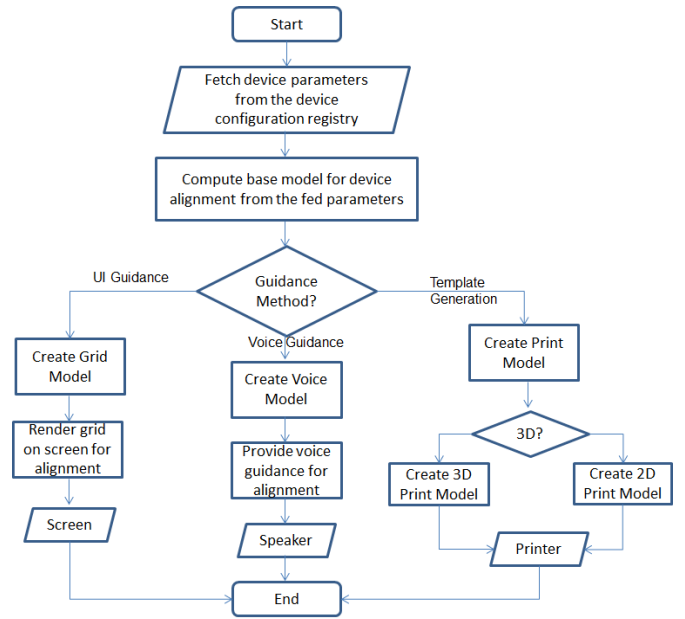

Fig. 4. Flowchart of the smartphone camera alignment process in the two device case, using visual (UI) guidance, voice guidance and template generation.

Fig. 4 shows a flowchart of the process in all the three cases: voice, visual and holder apparatus. The initial steps are common: the application would have to calculate the base model for alignment after inputting the dimensions of the devices and the positions of the cameras etc. Once the base model is computed, the alignment help can be given to the user in any of the three ways mentioned: visual, voice or holder template.

Fig. 5 illustrates the visual method to help the user align the two smartphones by displaying gridlines.

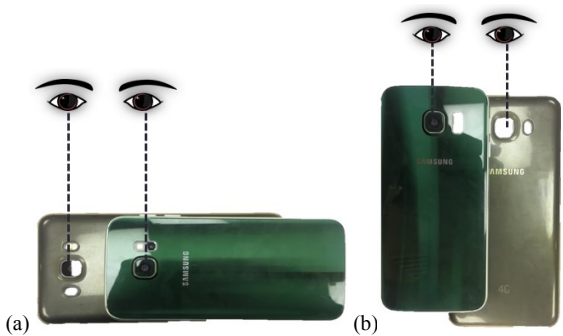

(a)  (b)

Fig. 3. Illustration of the alignment of two smartphones for 3D content capture in the (a)horizontal and (b) vertical case. In each case the distance between the smartphone cameras is kept equal to the inter pupillary distance of 6.5 cm.

Fig. 3 shows the aligned smartphones in the horizontal and vertical cases.

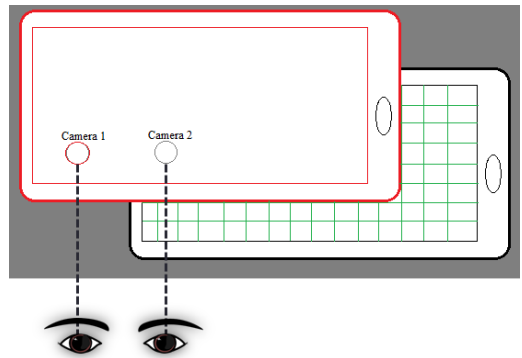

Fig. 5. Illustration for the system for alignment of two smartphones for 3D content capture, using two smartphones cameras along with an alignment grid.

## VII. GENERATION OF THE PRINTED HOLDER TEMPLATE

As mentioned, the holder template is generated based on the dimensions of the two mobile phones as well as the positions of their back cameras. Although both 2D cardboard template and 3D template are possible, for now we focus on the

2D cardboard template for simplicity. The idea is to have a holder where both the phones will be fitted, in such a way as to make it easier for the user to capture simultaneous videos with both the phones while keeping them aligned and their cameras separated by the inter-pupillary distance.

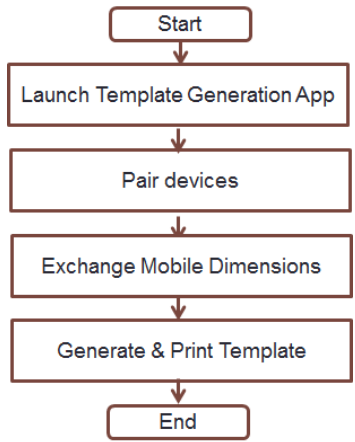

Fig. 6. Flowchart of the template generation process.

For generation of the template, the devices to be used are paired and the characteristics of the devices (such as device dimensions and locations of the cameras) are exchanged. Based on this information, the template for the apparatus can be generated and printed.

The dimensions and camera positions are obtained from the server, where they are stored, upon inputting the model numbers. Alternatively, the user can input them manually using our application. The template generation process flowchart is illustrated in Fig. 6. Similar to Google cardboard, we use simple Velcro strips to hold the straps in place. Our design is meant to be simple and functional. Our app takes as input the phone dimensions, and outputs a design, which the user can then attach to the cardboard and cut it out, fixing Velcro strips and then it is ready to use.

The diagram on fig. 7 illustrates the design of the cardboard template, for the case in which both the phones have identical dimensions and the back cameras are towards the center, close to the top edge of the phones. Other configurations for camera positions and phone dimensions etc. shall follow similar principles. For the heterogeneous dimensions case, where the two phones are of different sizes and/or back camera positions are different (say one is on the side and another at the center) we would have to calculate the strap and cardboard sizes in similar ways. The main thing to note is that the distance between the cameras should be equal to the inter-pupillary distance of 6.5 cm.

For the case considered, where the two phones are of the same dimensions and back cameras are at the center and near the top edge; our design includes four cardboard straps to hold both the phones in a stable way. The positioning and dimensions of the straps is shown in Fig. 7.

The length of the longer and horizontal cardboard straps (straps 1, 3 and 5) is given by:

$$V + (H+M) + W + H \qquad (1)$$

Where V is the length of the Velcro strip, H is the height (thickness) of the mobile phone, M is the cardboard thickness (added to account for the fold of the cardboard), W is the width of the phone. H (phone thickness) is repeated because there are two folds of the strap to hold the device in place. The longer straps hold the phones within the cardboard.

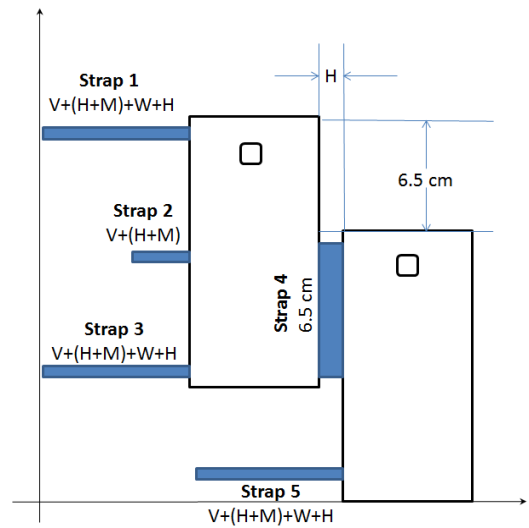

Fig. 7. Diagram showing the calculation of the lengths and positions of the straps for a cardboard holder to align two smartphones for VR content capture. We assume the dimensions of both the phones are identical. Here V is the length of the velcro strip, H the height (thickness) of the phones, M the cardboard material thickness, W the width of the phone. The model is then printed and a cardboard is cut out to the required dimensions, and the velcro strips attached. The four straps fitted with velcro ensure the two phones are held securely.

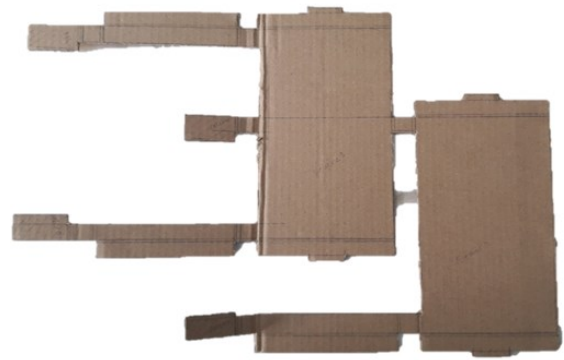

Fig. 8. Photo of the cut out cardboard template for aligning two Samsung J7 smartphones, such that their cameras are kept apart at the inter pupilliary distance of 6.5 cm.

The shorter strap 2 is of length

$$V + (H+M) \qquad (2)$$

Strap 4 is of width H and not 2H, since the phones are not folded face to face but rather back to back.

Fig. 8 shows a photo of the resultant cut out cardboard holder template for aligning two Samsung J7 smartphones, created as per the design in fig. 7.

VIII. CAPTURE OF VR CONTENT

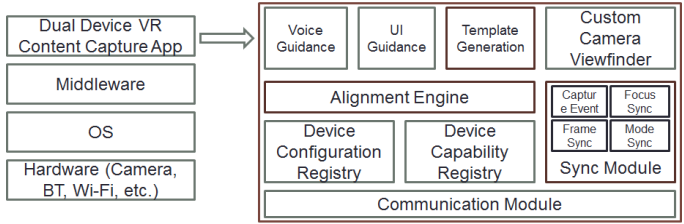

Fig. 9.  Block diagram of the system, showing the modules of the VR Content capture app for two smartphones.

For capturing the VR content, once the smartphones are aligned properly using our holder template or the other methods, we use the VR content capture app. Fig. 9 shows the block diagram of the system. We assume that both the phones have our VR capture application running on them to register and pair the camera apps for VR content generation. After the pairing process for the camera apps on the two smartphones is complete, the system running in the background fetches the device dimensional attributes and calibrates the UI accordingly to align the devices, taking into consideration factors such as camera capabilities, device resolution and dimensions.

When the user launches the dual device VR content capture application for the first time, the system automatically detects the model details of the phone and downloads the required device configuration and the compatible models. During the pairing process, the user is presented with a list of available compatible devices. Once the user has selected a particular device from the list, the system automatically determines the common denominator of the functionalities and configures the application. The configuration is stored in the device configuration registry, which is input to the alignment engine.

The system has registries comprising of device capabilities and configurations. This can be received in real time or can be pre-configured. Based on the device capabilities and configuration, the alignment engine determines the alignment for the devices and cameras. Once aligned and videos captured, the feeds can be easily merged. In case one phone camera has higher specifications such as frame rate and resolution, the lower of the two is used in both the devices to capture the video.

The main modules of the system are described below.

A.  Alignment Engine

The alignment engine is the key component of our VR content capture system. This makes use of the device dimensions and camera locations in the two devices to determine the alignment of the devices for VR content generation. It first identifies the device models and based on the given set of device models, queries the device configuration registry (hosted on the server) to get the device specifications like their dimensions and camera locations on the bezel. It also communicates with the device capability registry to find the capabilities that matches in its functionality in the respective devices for high quality media capture, including parameters such as resolution and frame rate.

This engine computes the base model for aligning the two devices based on the data fetched from device configuration registry. The base model contains:

- Co-ordinates of camera in "mm"
- Co-ordinates with which the second phone camera must be aligned.
- Co-ordinates of a box within which second phone must fit.
- Similar information for both the orientations of the phone.

Post configuration, the alignment engine prompts the user to select one of the below three device alignment guidance options:

B.  Device Configuration Registry

This gets the device configuration details either from the device or from the cloud server. It primarily includes device dimensions, camera locations and other details to evaluate the alignment.

C.  Device Capability Registry

This determines the highest common denominator of the functionalities of the two cameras to generate the VR content to the best possible quality.

D.  Sync Module

The sync module is another component that is used here, used to synchronize the two smartphone feeds and events. This includes the following four major elements:

- Capture Event: This ensures that both devices start capturing the scene at the same time.
- Focus Sync: This synchronizes the focus depth of the two devices.
- Frame Sync: This ensures that the (frames per second) fps rates of both the device cameras are in sync.
- Mode Sync: This synchronizes the camera capture modes (e.g. monochrome) of the two devices.

E.  Communication Module

The communication module is used to pair the devices and then enable them to communicate among each other for all the necessary capability exchange or handshake and all the functionality syncing.

The flowchart of the process is illustrated in Fig. 10. The devices to be used are paired and the characteristics of the devices (such as camera specifications) are exchanged. Based on this information, the 3D and/or VR content can be captured. All these functionalities require support from the camera module for getting and setting these parameters. We use camera module APIs to support the sync functionality of the device cameras (e.g. getFocusMode() and setFocusMode()). Once the phones are aligned and the video is captured using both the smartphones, VR content is rendered using a dual player system on the VR device or else standard stereoscopic techniques can be used to merge the frames to create VR content.

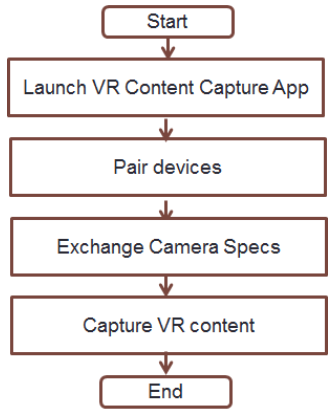

Fig. 10. Flowchart of the VR Content capture process, using the printed template and fitting in two smartphones.

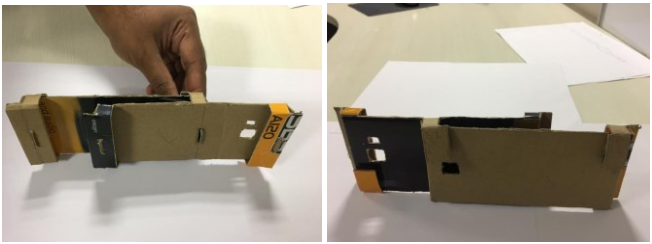

Fig. 11. Front and back view of the folded cardboard for aligned VR capture using two Samsung J7 smartphones, ensuring that the distance between the cameras matches with the inter pupillary distance.

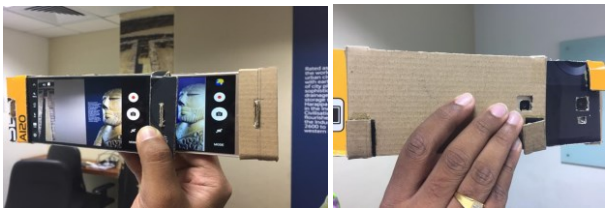

Fig. 12. Front and back view of a person holding the cardboard template with two Samsung J7 smartphones fitted in, while capturing a VR video.

## IX. PROTOTYPE IMPLEMENTATION DETAILS

We have designed and printed out a cardboard template prototype to align two Samsung J7 smartphones, and used the apparatus to capture a variety of 3D videos.

Fig. 11 shows the front and back view of the apparatus. Fig. 12 shows the apparatus with two mobile phones fitted in. As one can see, it is quite easy to hold both the phones with a single hand and capture the video, when fitted in the cardboard holder. For the purposes of getting a functioning proof of concept, we have used the Camerada app [12] to perform the merging of the video feeds.

## X. CONCLUSION AND FUTURE WORK

In this paper, we have presented a system to align smartphones in order to capture 3D or VR content. We have suggested solutions to generate content from one, two and three smartphones. We hope the system can help to align the devices and thus make it easier for users to capture the content. Eventually, this should help in increasing the popularity of VR. A patent for the system has also been filed in the India patent office [15].

In future we seek to extend the system for heterogeneous phones as well, where device dimensions and specs of the phone cameras are different.

REFERENCES

[1] Wikipedia. Interpupillary distance. [Online]. Available: https://en.wikipedia.org/wiki/Interpupillary_distance

[2] Gordon CC, Blackwell CL, et. al. Anthropometric Survey of US Army Personnel: Methods and Summary Statistics. Army Natick Soldier Research Development and Engineering Center , 2014.

[3] Richard Trenholm. The best 3D cameras for 3D photos. CNET, Mar 9 2011. [Online] Available: https://www.cnet.com/news/the-best-3d-cameras-for-3d-photos/

[4] Wikipedia. List of 3D enabled mobile phones [Online]. Available: https://en.wikipedia.org/wiki/List_of_3D-enabled_mobile_phones

[5] Vlad Savov. Dual-camera phones are the future of mobile photography. The Verge. Apr 11, 2016. [Online]. Available: https://www.theverge.com/2016/4/11/11406098/lg-g5-huawei-p9-two-camera-smartphone-trend-apple

[6] Han Jin. Lucidcam, the world's first consumer 3D VR camera. Indiegogo. [Online]. Available: https://www.indiegogo.com/projects/lucidcam-the-world-s-first-consumer-3d-vr-camera#/

[7] Lucidcam Virtual Reality 3D. [Online]. Available: https://www.lucidcam.com

[8] Samsung Gear 360 Specifications [Online]. Available: http://www.samsung.com/global/galaxy/gear-360/specs/

[9] Google. Get your cardboard. [Online]. Available: https://vr.google.com/cardboard/get-cardboard/

[10] Maria Korolov. 22 sites where you can buy Google Cardboard kits. Hypergrid Business. August 10, 2014. [Online]. Available: http://www.hypergridbusiness.com/2014/08/where-to-buy-google-cardboard/

[11] Scott Stein. Samsung Gear VR Review. CNET. 2017. [Online]. Available: https://www.cnet.com/products/samsung-gear-vr-2017/review/

[12] Camerada [Online]. Available: https://camerada.co

[13] Dean Takahashi. Camarada app lets you take virtual reality videos with two smartphones. Venturebeat, Jan 31, 2017 [Online]. Available: https://venturebeat.com/2017/01/31/camarada-app-lets-you-take-virtual-reality-videos-with-two-smartphones/

[14] N.D.Woodman et. al. Modular Configurable Camera System US Patent US20170054968, Owner GoPro. Filed Apr 2, 2012.

[15] Ramanujam RS, Dipin KP, Joy Bose. 3D and/or VR Content Capture. India patent No. 201741024884. Filed July 13, 2017